\documentclass[12pt,preprint]{aastex}

\lefthead{Brunetti et al.}
\righthead{Constraining Cosmic rays in galaxy clusters}

\begin{document}

\title{Cosmic rays and Radio Halos in galaxy clusters :
new constraints from radio observations}

\author{G. Brunetti\altaffilmark{1}, T. Venturi\altaffilmark{1},
D. Dallacasa\altaffilmark{1,2},
R. Cassano\altaffilmark{1,2},
K. Dolag\altaffilmark{3}, S. Giacintucci\altaffilmark{1,4},
G. Setti\altaffilmark{1,2}}

\altaffiltext{1}{INAF/IRA, via Gobetti 101,
I--40129 Bologna, Italy}
\altaffiltext{2}{Dip. Astronomia, University of Bologna, via Ranzani 1,
I--40127 Bologna, Italy}
\altaffiltext{3}{Max-Planck-Institut f\"ur Astrophysik, Karl-Schwarzschild
Strasse 1, D-85741 Garching bei M\"unchen, Germany}
\altaffiltext{4}{Harvard-Smithsonian Center for Astrophysics, 60 Garden
Street, Cambridge, MA 02138}

\begin{abstract}
Clusters of galaxies are sites of acceleration of charged particles and  
sources of non-thermal radiation. 
We report on new constraints on the population
of cosmic rays in the Intra Cluster Medium (ICM) 
obtained via radio observations of a fairly
large sample of massive, X--ray luminous, galaxy clusters 
in the redshift interval 0.2--0.4.
The bulk of the observed galaxy clusters does not show any hint of 
Mpc scale synchrotron radio emission at the cluster center
(Radio Halo).
We obtained solid upper limits to the diffuse radio emission 
and discuss their implications 
for the models for the origin of Radio Halos.
Our measurements allow us to derive also a limit to the 
content of cosmic ray protons in the ICM.
Assuming spectral indices of these protons 
$\delta =2.1-2.4$ and 
$\mu$G level magnetic fields, as from Rotation Measures,
these limits are one order of magnitude deeper
than present EGRET upper limits, while they are less stringent for
steeper spectra and lower magnetic fields.
\end{abstract}


\keywords{particle acceleration - radiation mechanisms:
non--thermal - galaxies: clusters: general -
radio continuum: general - X--rays: general}

\section{Introduction}

Clusters of galaxies are ideal astrophysical environments 
for particle acceleration.
Large scale shocks which form during the process of cluster formation
are believed to be efficient particle accelerators (e.g. Sarazin 1999;
Gabici \& Blasi 2003; Ryu et al. 2003; Pfrommer et al. 2006).
Cosmic rays (CRs) can also be injected into 
the ICM from ordinary galaxies and AGNs (e.g. V\"olk \& Atoyan 1999)
and turbulent eddies may contribute to the particle acceleration
process (e.g. Brunetti \& Lazarian 2007).
CRs accelerated within the cluster volume 
would then be confined for cosmological times 
and the bulk of their energy is expected in
protons since they 
have radiative and collisional life--times
much longer than those of the electrons (e.g. Blasi et al. 2007, for
a review).

While present gamma ray observations can only provide
upper limits to the average energy density of
CR protons in the ICM (e.g. Reimer et al. 2004), 
the presence of relativistic electrons in a number of clusters
has been ascertained via the detection of 
a tenuous synchrotron radio
emission: giant Radio Halos (RHs)
and mini-Radio Halos, fairly symmetric sources at the cluster
center, and 
Radio Relics, elongated 
sources at the cluster periphery (e.g. Feretti \& Giovannini 2007).
It is customary to classify the models for the origin
of RHs in 
{\it secondary electron} (e.g. Blasi \& Colafrancesco 1999)
and {\it reacceleration models} (e.g. Brunetti et al. 2001;
Petrosian 2001),
depending on whether the radiating electrons are 
produced as secondary
products of hadronic interactions or 
reaccelerated by turbulence from a
pre-existing population of non-thermal seeds in the ICM, 
respectively.
These models predict a different connection between
radio and X-ray properties of clusters which are discussed
in this Letter and compared with new observations:
in Sect.2 we review the expectations of the different models,
in Sect.3 we briefly present the radio observations of
our cluster sample, and
in Sects.4 \& 5 we report and discuss our results.
Concordance ($H_o$=70, $\Omega_m$=0.3, $\Omega_{\Lambda}$=0.7)
cosmology is used.

\section{Radio -- X-ray correlation \& origin of RH}

Giant RHs follow a correlation
between their radio power at 1.4 GHz
($P_{1.4}$) and physical size, and the X-ray luminosity ($L_X$)
and temperature of clusters in which they are found 
(e.g. Liang et al. 2000;
Bacchi et al. 2003; Cassano et al. 2006,07).

In this Letter we focus on the $P_{1.4}$--$L_X$ correlation 
that relates directly observable quantities.
The bulk of giant RHs has been discovered from the analysis
of relatively shallow surveys (NVSS, Giovannini et al.~1999, G99;
WENSS, Kempner \& Sarazin 2001, KS01) and a relevant issue is how 
observational biases may affect this correlation.
There is agreement on the fact that the upper envelope 
of the $P_{1.4}$--$L_X$ correlation
is likely to be solid (e.g. Clarke 2005), but 
the effect of observational biases on the 
lower envelope is more problematic
(Rudnick et al. 2006).
Indeed, if all clusters would have cluster--scale radio emission 
at the level of presently known RHs, then the $P_{1.4}$--$L_X$
trend may possess a fairly large spread with lower luminosity RHs
falling just below present observational limits.
On the other hand it is also possible that clusters have a physical
bimodal distribution with the 
RH--clusters following the correlation and
with other clusters having no (or much weaker)
cluster--scale radio emission.

The first possibility is essentially
what {\it secondary} models would expect (e.g. Miniati et al. 2001;
Dolag \& Ensslin 2000).
This comes from the combination of two points: 
the magnetization at $\mu$G level
is believed to be a very common property of the ICM (Clarke et al. 2001;
Govoni \& Feretti 2004),
and CR protons accumulated in galaxy clusters
for cosmological time scales provide a fairly stable 
continuous source of injection
of secondary electrons and positrons in the ICM.

On the other hand, an unavoidable prediction of the
{\it re-acceleration} scenario is a bi-modality of clusters.
In this scenario particles are supposed to be re--accelerated by 
MHD turbulence in the ICM and this requires enough turbulence to boost 
electrons at the energies necessary to emit synchrotron radiation
at GHz frequencies. Thus 
giant RHs should be strictly connected to massive and merging systems 
where indeed enough turbulence can be developed 
(Cassano \& Brunetti 2005), and should live for relatively 
short time--scales (1 Gyr or less) because of the finite dissipation 
time--scale of turbulence.
In this scenario merging clusters are expected to move with time
from a {\it radio quiet} region in the $P_{1.4}$--$L_X$ plane 
to the $P_{1.4}$--$L_X$ correlation. This happens in a relatively
short time--scale, of the order
of $\approx 10^8$ yrs (i.e. the acceleration 
time--scale of the emitting particles, see Fig.19 in Brunetti et al.~2004), 
and thus the region between
RHs and {\it radio quiet} clusters in the $P_{1.4}$--$L_X$ plane
should be poorly populated.

Thus, the distribution of clusters in the
$P_{1.4}$--$L_X$ plane is 
important to constrain current models.

\section{Cluster sample and GMRT observations}

From the ROSAT--ESO Flux Limited X--ray (REFLEX) galaxy cluster catalog
(B\"ohringer et al. 2004) and from the extended ROSAT Brightest
Cluster Sample (eBCS) catalog (Ebeling et al. 1998, 2000) we selected
all clusters with $0.2\leq z \leq 0.4$,
L$_{\rm X}$(0.1--2.4 keV) $>$ 5 $\times$ 10$^{44}$ erg s$^{-1}$,
and with declination $\delta\geq-30^{\circ}$ (for the REFLEX) and
$15^{\circ}<\delta<60^{\circ}$ (for the eBCS).
These selection criteria led to a sample of 50 X--ray selected
galaxy clusters (27 from REFLEX, 23 from eBCS) 
with similar luminosity. The sample includes 6 clusters
with well studied RHs (A\,2744, A\,1300,
A\, 2163, A\,773, A\,2219, A\,2390, e.g. Feretti \& Giovannini
2007; Bacchi et al. 2003) and A\,1758 for which hint of diffuse
emission is also reported (G99, KS01).

We carried out GMRT (Giant Metrewave Radio Telescope) observations at
610 MHz only 
for 34 clusters in the sample (those with no high sensitivity radio
information already available, and not included in the GMRT 
cluster key project); each cluster was observed for 2--3 hrs (hour angle
3--5). Thanks to the dirtibution of antennas at GMRT we 
obtained images for each cluster with resolutions
ranging from $\sim 6^{\prime\prime}$ to $\sim 25^{\prime\prime}$ and
r.m.s. (1$\sigma$)$~\sim 30 - 180~\mu$Jy/b (Venturi et al.
2007, V07; Venturi et al., in prep) which allows us to image both compact and
extended sources in the fields.
We detected RHs in 4 of them (RXCJ\,2003--2323; A\,209
and RXCJ\,1314--2515, V07; A\,697 Venturi et
al. in prep.; a Relic was also found in A\,521, Giacintucci et
al. 2006).
No hint of cluster--scale
radio emission was found in the remaining 29 clusters.
For these clusters it is necessary to place solid upper limits to the flux
density of their Mpc scale radio emission (few arcmin at the
redshifts of our clusters).
In Fig.~1 we report the normalised integrated brightness profiles
of well studied RHs: they are quite similar
and $\approx$50\% of the luminosity,
$L_H$, is emitted within about half radius, $R_H$. 
Detection limit based on the brightness within $R_H/2$ gives
a 610 MHz luminosity in W/Hz ($\theta_b$ is the beam-FWHM):

\begin{equation}
L_H \geq 3.5 \cdot 10^{23} \big(1+{{z}\over{0.25}}\big)^4 
{{ (r.m.s.)} \over{
50 \mu Jy/b}} \big( {{25^{\prime \prime}}\over{\theta_b}} \big)^2
\big( {{R_H }\over{0.5 Mpc}} \big)^2
\label{Lmin}
\end{equation}

\noindent
which is $\approx$50 times smaller than luminosities 
of known giant RHs at $z \approx 0.25$ (V07).

In order to derive more solid constraints to use in this paper
we inject fake RHs in our datasets.
We model the brightness profile in Fig.~1 with sets of
optically thin spheres with different radius and flux densities,
and obtain {\it families} of fake RHs with total flux
densities S$_{RH}$ ranging from 3 to 300 mJy and angular diameters
from 180 to 350 arcsec. 
Those fake RHs were injected into the uv--components of 
our cluster datasets by means of the task UVMOD in AIPS, 
and the resulting
datasets were imaged with the procedures given in V07
with the task IMAGR at resolutions in the range 
10--25 arcsec. 
The injected flux density of RHs is not fully recovered by the imaging
and an increasing fraction of injected flux is lost when S$_{RH}$
decreases, and/or the total angular size increases; an example
is given in Fig.~2.
We also found very little dependence on the resolution of the image, 
at least in the range 15$^{\prime\prime}-25^{\prime\prime}$. 
We find that 
the lowest value of the injected flux density that leaves a residual
flux in the images which can be reasonably interpreted
as due to an extended low brightness radio source on the basis of the
standard radio imaging is in the range $\approx$5--12 mJy.
This marks the value of the upper limit to the injected flux 
of RHs and scales both with the largest angular size of the fake RHs
and with the r.m.s of the final image (e.g., Fig.~3). Note however,
that at this point 
a residual flux is still formally detected at 4--5$\sigma$ level 
on an area of a few beams in the low resolution images (e.g., Fig.~2).
Our limits should thus be considered as conservative; they are 
typically $\approx$2.5 times larger than those
in Eq.(\ref{Lmin}).
\\
We derive these solid limits to the detection of RHs 
for 20 clusters observed at the GMRT.
Indeed, among the observed 
29 clusters with no hint of cluster--scale emission, we excluded 
A 3444, A 1682 and Z 7160 where extended radio galaxies in the field
makes difficult to straightaway apply our procedure, and also excluded 
6 clusters in our sample with poor 
quality of the data (r.m.s. $> 120 \mu$Jy$/$b caused by 
interferences).

\section{Results}

In Fig.~4 we report the distribution of clusters 
in the $P_{1.4}$--$L_X$ plane.
Giant RHs and upper limits obtained from UVMOD simulations (Sect.~3)
with $R_H =$0.5 Mpc are reported in magenta. 
These solid upper limits lie about one order of magnitude below 
the correlation for giant RHs. 
This allows us to firmly establish that the $P_{1.4}$--$L_X$
correlation (solid line in Fig.~4, from Cassano et al.2006)
is real and that its lower envelope is not driven by
observational biases
(at least for $L_X \geq 5 \times 10^{44}$erg s$^{-1}$,
i.e. the selection limit of our clusters).

Most importantly, we find that clusters with similar $L_X$ and 
redshift have a clear bimodal distribution.
Cluster-scale radio emission
at the level of presently studied RHs is not ubiquitous in galaxy
clusters and only $\approx 1/3$ of clusters in our 
sample host a RH. Although no homogeneous high resolution
X--ray data are still available for all our clusters, 
the RHs are found in dynamically disturbed
systems, while clusters without RHs are either disturbed or
relaxed systems (V07).
Fig.~4 (green arrows) shows that 
even more stringent upper limits are obtained considering radio
emission on cluster--core scale ($R_H$=0.25 Mpc).

It is challenging for {\it secondary} models to accommodate 
the observational picture of Fig.~4. 
These models would expect the cluster--scale radio
emission much common and predict
some general $P_{1.4}$--$L_X$ trend for all clusters 
with some scattering due to 
the effect of the different
CR proton content and magnetic field among clusters
(e.g. Miniati et al. 2001; Dolag 2006).
Thus strong dissipation of the magnetic field in the ICM
in relatively short time-scales is necessary 
to reconcile {\it secondary} model expectations with the data, 
although reconciling this dissipation with present theoretical
understanding (e.g. Subramanian et al. 2006) and data 
(e.g. Govoni \& Feretti 2004) might be problematic.
Recently Pfrommer (2007) presented numerical simulations
of {\it secondary} \& {\it shock accelerated} 
particles. Also in this case extended synchroton emission, at least
on cluster--core scale, is predicted to be common, at the level
of presently known RHs, and clusters are predicted to  
follow relatively well defined correlations (Fig.1 in
Pfrommer 2007) with 
merging and non merging systems lying on the upper and lower envelope 
of correlations, respectively. Thus similar considerations can be
applied also to this scenario.

\section{Limits on CR in non-radio emitting clusters}

Gamma ray observations of a number of nearby galaxy clusters
limit the energy density
of CR protons in these clusters to 10--20 \% 
of the thermal energy (Pfrommer \& Ensslin 2004; Reimer 2004).

The upper limits for clusters with no RHs of our sample
allow us to obtain indirect upper limits to the energy 
density of CR protons
in these clusters. Indeed, by requiring that 
radio emission from {\it secondary} e$^{\pm}$ is 
below the upper limits in Fig.~4 one gets 
constraints on the content of CR protons from which secondaries
are injected.

We use the formalism in Brunetti \& Blasi (2005) to
calculate the stationary spectrum of secondary pairs 
in the ICM.
Typical limits to the CRs content 
are given in Fig.5 by simply assuming average values of thermal density
($n_{th}=10^{-3}$cm$^{-3}$) 
and magnetic field in a sphere of 
radius $R_H$=0.5 Mpc: 
for 
$\geq \mu$G fields and relatively flat spectra of the CRs
our limits are about one order of magnitude 
deeper than present EGRET upper limits; note that our limits
scale with $(n_{th}/10^{-3})^{-2}$.
Limits are also significantly 
lower than the typical CR energy content of clusters
that is claimed from numerical simulations in which CRs are 
accelerated at large scale shocks (e.g. Ryu et al. 2003).
\\
On the other hand, for steeper CR spectra (or lower values of the field)
the synchrotron constraints 
become gradually less stringent and the energy content of 
CRs in our clusters may be considerably larger.

\section{Conclusions}

We have reported on constraints on the  
origin of RHs and on the CR content in the ICM obtained via radio
observations of a fairly large sample of X--ray luminous clusters
at $z = 0.2-0.4$.

\noindent
In the bulk of these clusters we do not find evidence
of Mpc--scale radio emission at the level of RHs. 
Our conclusions become even more stringent 
considering radio emission on cluster--core scale, typical of smaller 
RHs and mini-Halos.

\noindent
We firmly confirm that RH--clusters 
follow a {\it physical} correlation 
between synchrotron and X--ray luminosities.
We find that clusters have a bimodal distribution 
in the $P_{1.4}$--$L_X$ plane (Fig.~4); this is in line
with the expectation of the {\it re-acceleration scenario}.
On the other hand, in order to reconcile these observations 
with expectations from {\it secondary} models 
strong dissipation of the magnetic field in 
the clusters with no radio emission is necessary.

\noindent
Our measurements allow us to also derive simple limits on the presence of
CR protons in the ICM (Fig.~5). 
In the case of relatively flat spectral energy 
distribution of these CRs stringent upper limits can be obtained:
the energy density of CRs should be $\leq 1$\% of the thermal energy 
in case of $\geq \mu$G field strength. This would make problematic
the detection of gamma rays from
$\pi^o$--decay in clusters with GLAST.
On the other hand, by assuming steeper spectral energy distributions
of these CRs (or lower magnetic fields)
our limits become less stringent. 

\section{Acknowledgements}
We acknowledge M.Bondi and an anonymous referee for useful comments, and
support through grants ASI-INAF I/088/06/0 and PRIN-MUR 2006-02-5203.

\clearpage

\begin{figure}
\plotone{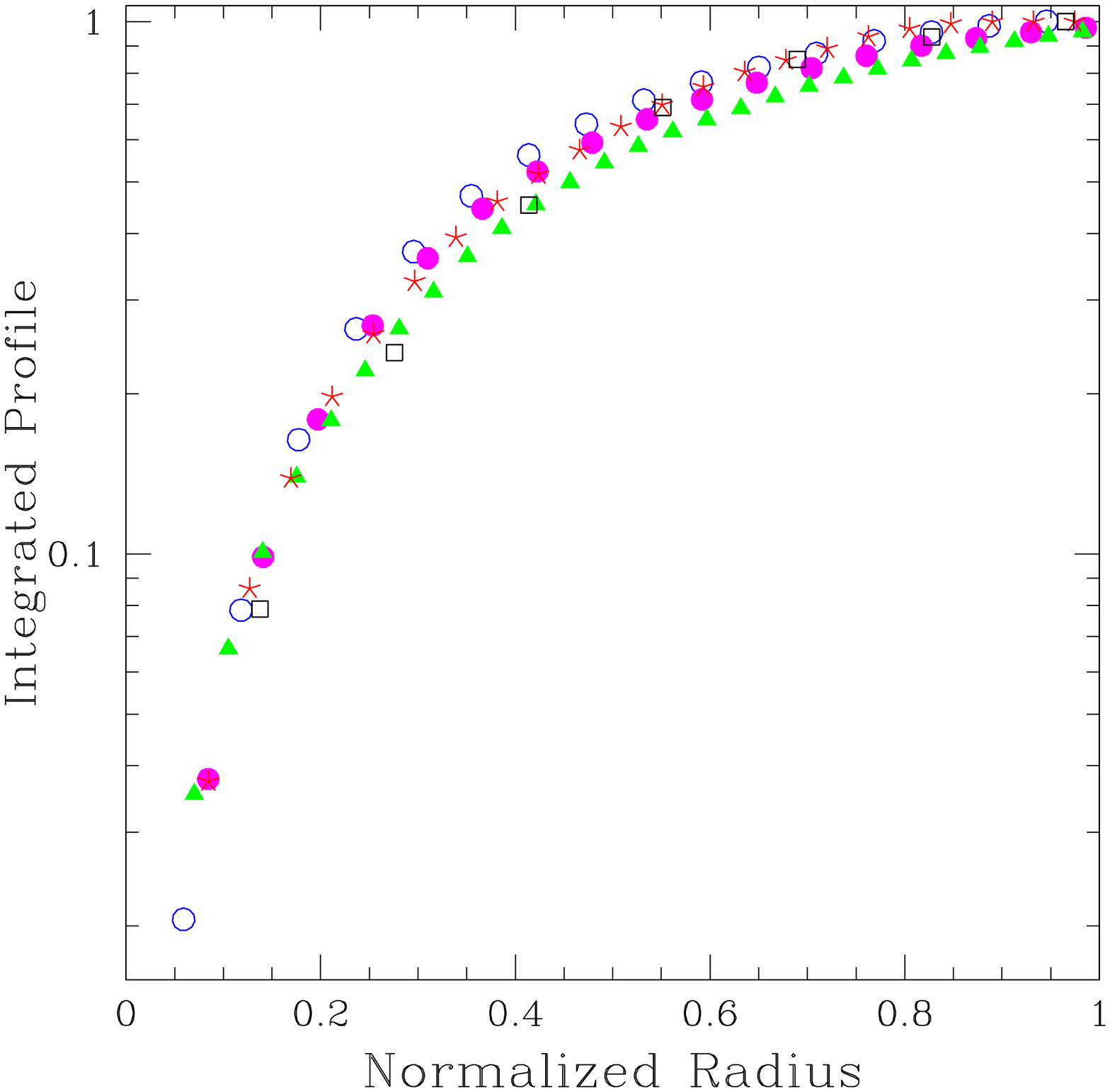}
\caption[]{Integrated brightness profiles of well studied
RHs (from Cassano et al. 2007): A545 (boxes), A2319 (triangles), 
A2744 (empty circles), A2163 (filled circles), A2255 (stars).
}
\label{fig:profili}
\end{figure}

\begin{figure*}
\plotone{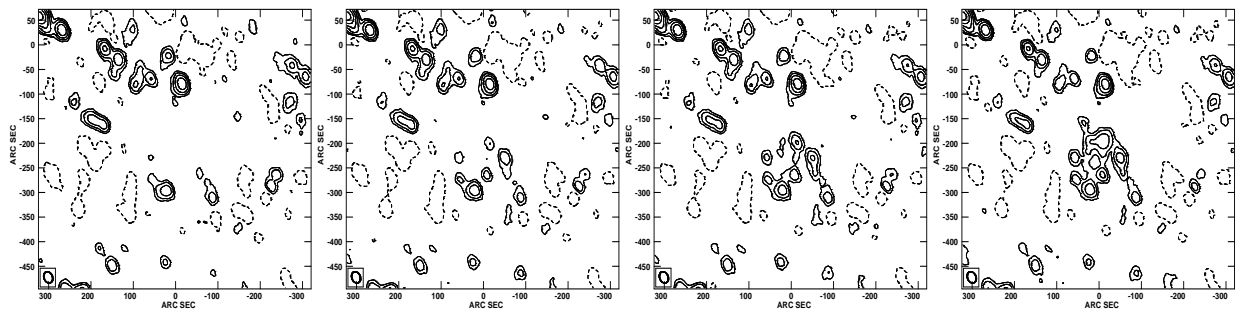}
\caption[]{Example of injection of fake RHs with apparent radius
$\Theta_H = 150$ arcsec and $S_{RH}=$ 0, 8, 11 and 15 mJy (from left to 
right). The r.m.s. of the image is 65 $\mu$Jy/b ($\theta_b 
\simeq 20 \times 24$ arcsec).
Contour levels are given for 0.1$\times$(-1, 1, 2, 4, 8, 16, 32, 64)
mJy/b.
In this case diffuse emission is revealed with standard analysis 
(including comparison between high and low resolution images)
for $S_{RH} > $11 mJy.}
\label{fig:mappe}
\end{figure*}

\begin{figure}
\plotone{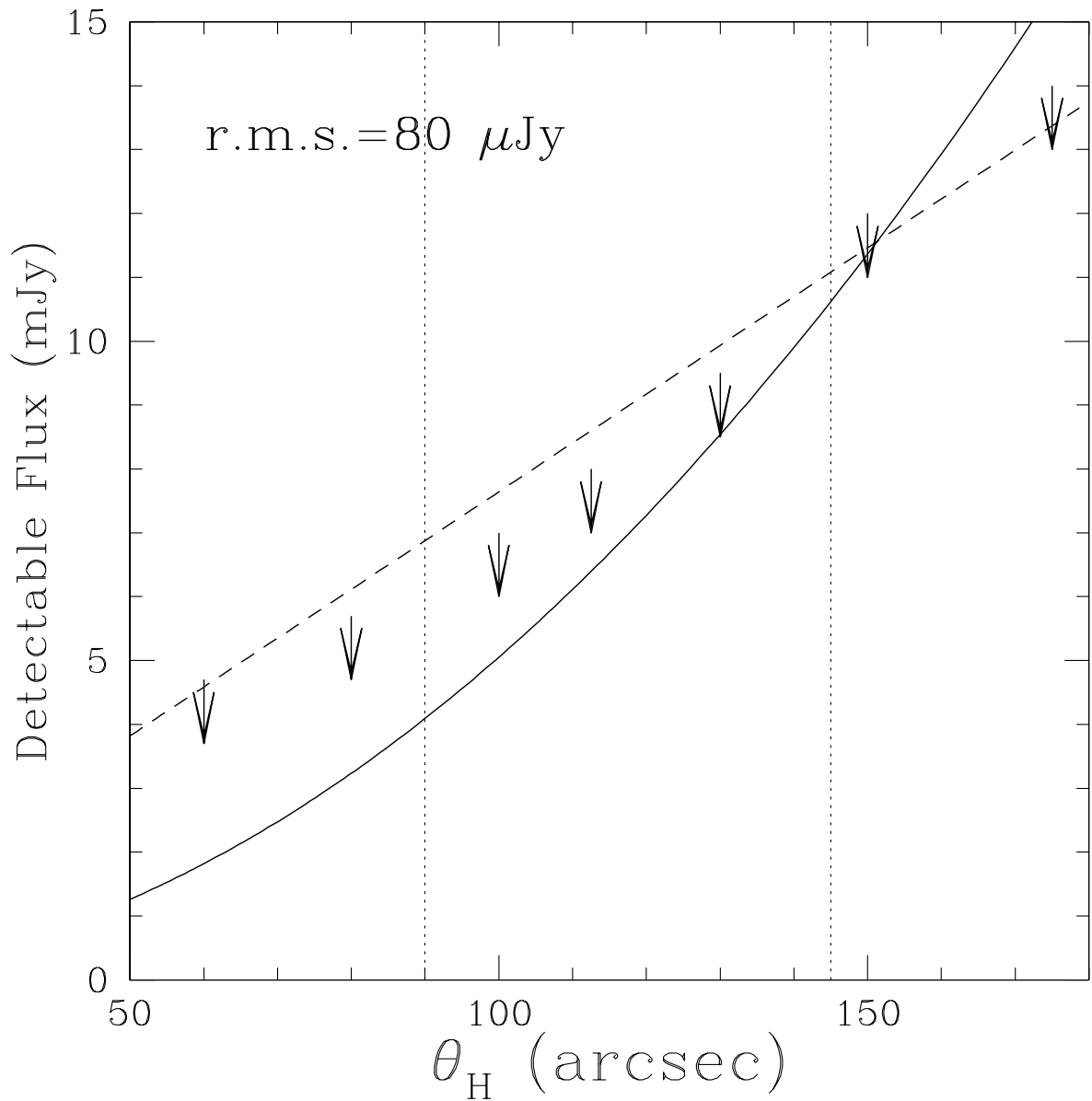}
\caption[]{Upper limits on the detectable RH flux (arrows) 
from UVMOD simulations (GMRT cluster
field with r.m.s.= 80 $\mu$Jy/beam and beam=20$\times$22 arcsec) 
as a function of apparent radius $\Theta_H$ (in arcsec).
The solid line marks the constant brightness scaling, the
dashed line marks the $1/\Theta_H$ scaling.
The vertical dotted lines mark the range of $\Theta_H$ 
spanned by our clusters ($R_H$= 0.5 Mpc).
}
\label{fig:ul_rh}
\end{figure}

\begin{figure}
\plotone{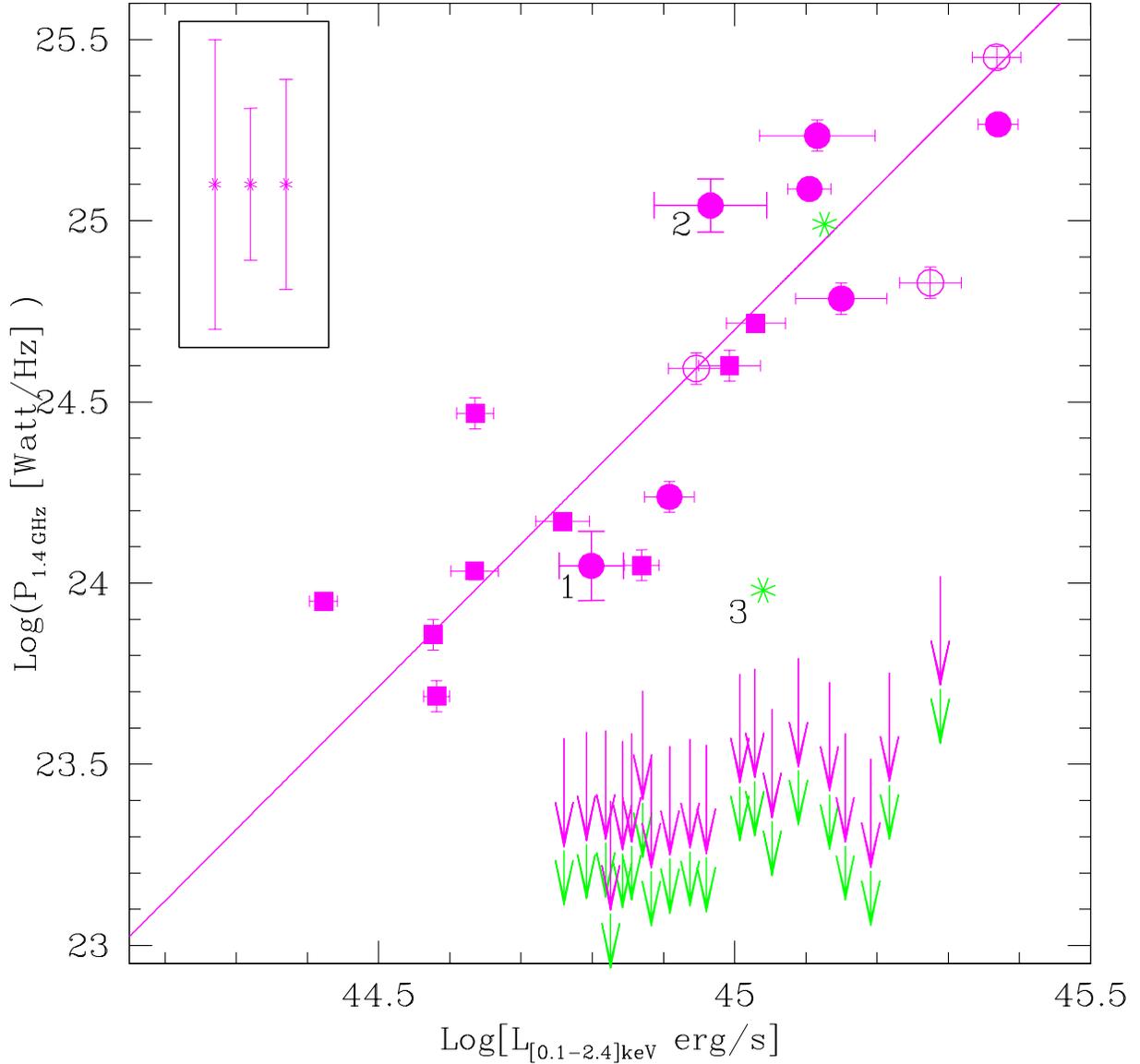}
\caption[]{Clusters in the $P_{1.4}$--$L_X$ diagram:
published giant RHs in our sample (magenta filled circles), 
other giant RHs at $z > 0.2$ (magenta empty circles) and at $z < 0.2$
(magenta filled squares). 
Upper limits are obtained assuming $R_H$=0.5 (magenta)
and 0.25 Mpc (green) and are scaled at 1.4 GHz with a typical
spectral index of RHs, $\alpha = 1.3$ 
(Feretti et al.~2004; $P(\nu)\propto \nu^{-\alpha}$).
The mini-Halo A2390 and the small RH RXJ1314
which are in our sample are also reported 
(green asterisks). RHs taken from GMRT observations are:
1= A209, 2= RXJ2003, 3=RXJ1314 (TV07).
Estimated dispersions in $P_{1.4}$ at fixed cluster mass/temperature
from simulations of {\it secondary} models 
are reported (from left to right: 
Miniati et al. 2001, Dolag \& Ensslin 2000, Pfrommer 2007).}
\label{fig:limiti}
\end{figure}

\begin{figure}
\plotone{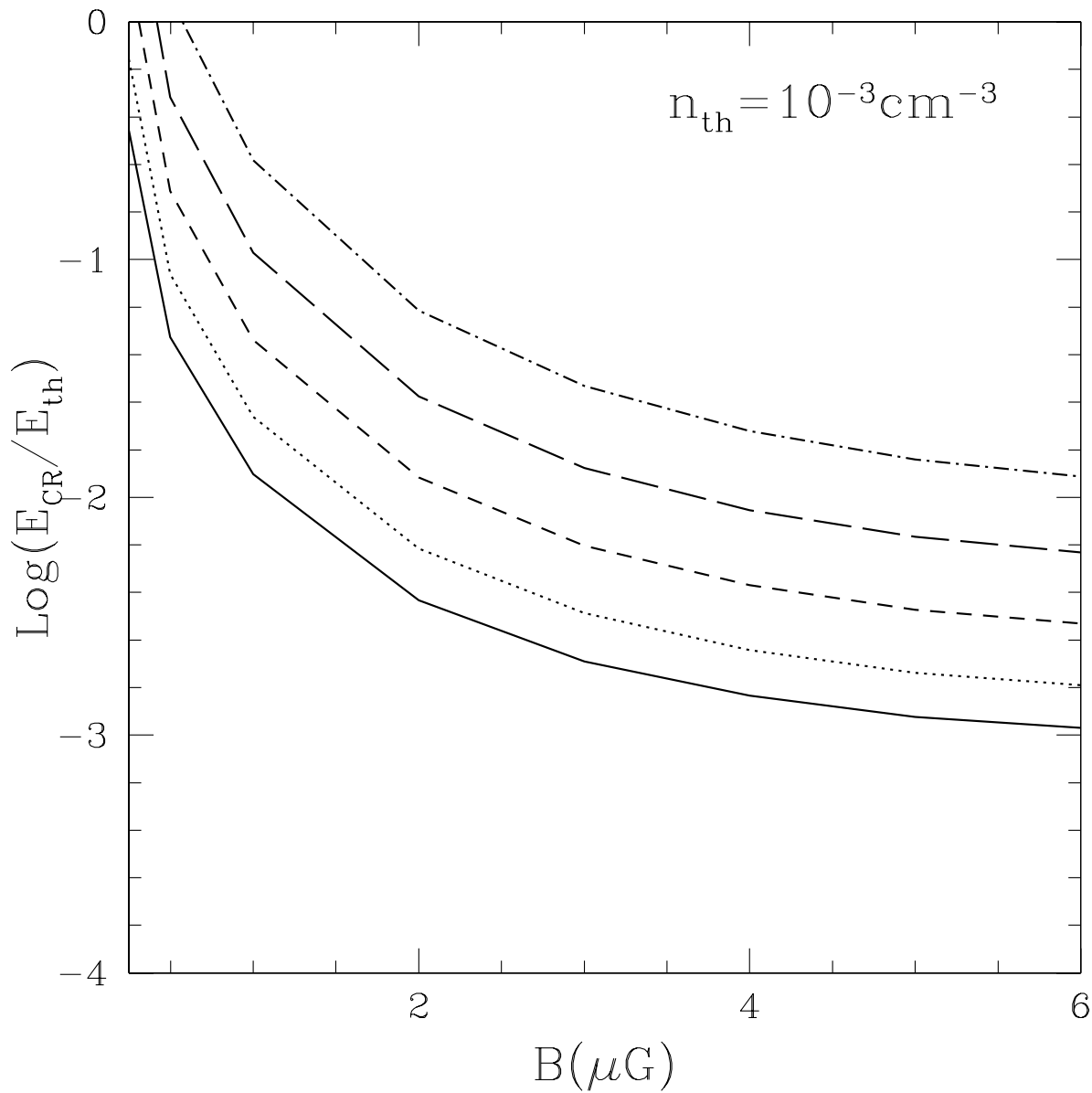}
\caption[]{Upper limits (curves) to the energy density 
(for $p >$0.01$m_p c$) of CRs (normalised to the thermal 
energy density) as a function of the magnetic field. 
We assume a CRs spectrum $\propto p^{-\delta}$
($\delta$=2.1, 2.3, 2.5, 2.7, 2.9 from bottom to top), 
$z=0.25$ (for inverse Compton losses), thermal density$=10^{-3}$cm$^{-3}$,
temperature$=10^8$K and synchrotron upper limits$=10^{24}$W/Hz
at 610 MHz.}
\label{fig:f_lim}
\end{figure}

\end{document}